\documentclass[a4paper,fleqn,usenatbib]{mnras}
\usepackage{amsmath,amssymb}

\usepackage{txfonts}

\usepackage[T1]{fontenc}
\usepackage{ae,aecompl}

\usepackage{graphicx}	
\usepackage{amsmath}	
\usepackage{amssymb}	

\title[Gravitational collapse in turbulent medium]{Criteria for gravitational
instability and quasi-isolated gravitational collapse in turbulent medium }

\author[Guang-Xing Li]{
Guang-Xing Li$^{1}$\thanks{E-mail: gxli@usm.lmu.de (USM)}
\\
$^{1}$University Observatory Munich, Scheinerstrasse 1, D-81679 M\"unchen,
Germany\\}

\date{Accepted XXX. Received YYY; in original form ZZZ}

\pubyear{2015}

\begin{document}
\label{firstpage}
\pagerange{\pageref{firstpage}--\pageref{lastpage}}
\maketitle

\begin{abstract}
 We study the evolution of structures in turbulent, self-gravitating media,
 and present an analytical criterion $M_{\rm crit} \approx \epsilon_{\rm cascade}^{2/3} \eta^{-2/3} G^{-1}
l^{5/3}$ (where $M_{\rm crit}$ is the critical mass,  $l$ is the scale,
$\epsilon_{\rm cascade}\approx \eta \sigma_{\rm v}^3 /l $ is the turbulence
energy dissipation rate of the ambient medium, $G$ is the gravitational
constant, $\sigma_{\rm v}$ is the velocity dispersion, $l$ is the scale and
$\eta\approx 0.2$ is an efficiency parameter) for an object to undergo quasi-isolated gravitational collapse.
The criterion also defines the critical scale ($l_{\rm crit} \approx \epsilon_{\rm cascade}^{1/2} \eta^{-1/2} G^{-3/4}
\rho^{-3/4}$) for turbulent gravitational
instability to develop.
The analytical formalism explains the size dependence of the masses of the
progenitors of star clusters ($M_{\rm cluster} \sim R_{\rm cluster}^{1.67}$) in our Galaxy.
\end{abstract}



\section{Introduction}
Astrophysical fluid systems are characterised by large sizes and long evolution
times. The Reynolds number, $Re =  U L / \nu = L^2 / T  \nu$ (where $U$ is
velocity, $L$ is scale, $T$ is time and $\nu$ is the viscosity) is typically
large.
Gravity drives
the formation of structures. One thus needs to understand the interplay between
the two.  We consider an object embedded in a turbulent ambient flow, and are
concerned with this question:
{\it Under what conditions can an object be considered ``detached'' from the
ambient medium, such that its evolution is ``quasi-isolated''? What is the
appropriate condition for gravitational instability to develop?}

One possible criterion is the virial parameter
\citep{1992ApJ...395..140B}, which quantifies the relative importance of gravity and turbulence in a given structure. However, one
 limitation of the virial parameter in its basic form is that it neglects the
 dynamical interaction between the structure and the ambient environment
\citep{2006MNRAS.372..443B}.
Turbulence is a process where energy has been transferred from larger to
smaller scales.
Since this energy transfer has not been explicitly considered, the virial
parameter in its basic form can not be used to study the interaction between the
 object and the ambient environment.

We derive an
analytical criterion for an object to be considered as quasi-isolated in a
turbulent flow. The criterion is derived by explicitly considering the interplay
between turbulence and gravity at the boundary of the object.
 We apply the criterion to the evolution of structures in the turbulent molecular interstellar
medium, and find that the observed properties of clumps hosting proto-star
clusters can be explained by our formalism. The criterion can be used to
study the development of gravitational instability in a turbulent medium
\citep{1951RSPSA.210...26C,1952Natur.170.1030P}.


\section{Overall picture}

\begin{figure}
\includegraphics[width = 0.5 \textwidth]{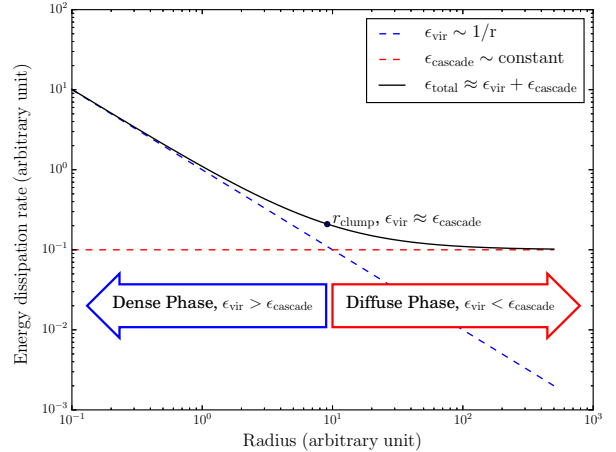}
\caption{ An illustration of the proposed picture. The object
has a density profile of $\rho/\rho_0 = (r / r_0)^{-2}$ (such that the energy
dissipate rate of the viralised turbulence is $\epsilon_{\rm vir} =
\sigma_{\rm v}^3 / r = (4 \pi G \rho_0)^{3/2} r_0^3 / r \sim 1/r$).
We use the blue dashed line to represent the energy disputation rate of the
viralised turbulence. Due to turbulence cascade, the ambient medium has an
almost-uniform energy dissipation rate of $\epsilon_{\rm cascade}$. It is
represented by the red dashed line.
The total energy dissipation rate of the system is approximated as
$\epsilon_{\rm tot} \approx  \epsilon_{\rm vir} + \epsilon_{\rm cascade}$, and is represented
by the black solid line. The boundary of the object is determined as the radius
where $\epsilon_{\rm vir} \approx \epsilon_{\rm cascade} $. Gas in the dense
phase satisfies $\epsilon_{\rm vir}  >\epsilon_{\rm cascade} $ and gas in the
diffuse phase satisfies $\epsilon_{\rm vir} <\epsilon_{\rm cascade} $.
The figure is created for illustrative purposes, and the normalisations of both
axes are arbitrary.
\label{fig:1}}

\end{figure}

In our picture, the medium is composed of two phases: in the dense phase,
gravity determines the level of turbulent motion, and in the diffuse phase, the
level of turbulent motion is almost universal. Our ``object'' is composed of gas
in the dense phase, surrounded by the ambient medium that belongs to the diffuse
phase. The gas in the dense phase is ``quasi-isolated'' in the sense that the
ambient medium is not able to influence its evolution significantly. 
 We assume that turbulence in the dense phase is
viralised, and turbulence in the ambient medium is characterised by a constant
energy dissipation rate $\epsilon_{\rm turb}$. This illustrated in Fig.
\ref{fig:1}.

In the case of the Jeans instability \citep{1902RSPTA.199....1J}, one is mainly
concerned with the interplay between gravity and thermal support in terms of
pressure -- the  instability occurs when the internal gas pressure is not
sufficient to prevent gravitational collapse. In the case where the ambient
medium is turbulent, one needs to consider a different picture.

In our proposed picture, turbulence can provide internal support against
gravitational collapse. However, this type of support must be distinguished from e.g. support from thermal
motion. Unless cooling is extremely efficient, the support from thermal motion
does not need to be sustained by an external energy source, and for thermal
support to be effective, one only  need it to satisfy the pressure equilibrium
$p_{\rm thermal} \approx p_{\rm gravity}$. When an object collapses, $p_{\rm
thermal} < p_{\rm gravity}$. Support from turbulence has a different
nature, in that turbulent motion does not sustain by itself. Turbulent motion is dissipative, and without energy injection, turbulence would decay within a few crossing times.
Thus, it is necessary to sustain the turbulent motion for it
to be effective in supporting against collapse. 

When an turbulence-dominated object is collapsing, it is often not the case
that the turbulent pressure is much lower than the pressure required to
support against gravity. For a system where the turbulence is viralised
(such that $\sigma_{\rm v}^2\approx G m / r$, $m$ is the mass and $r$ is the
size of the object, $\sigma_{\rm v}$ is the velocity dispersion), the
ram-pressure of the internal turbulent motion ($p_{\rm turb} \approx \rho
\sigma_{\rm v}^2$, where $\rho$ is the density) is always comparable to the
pressure from self-gravity ($p_{\rm gravity} \sim G m^2 / r^4 \sim G \rho^2
r^2$). However, turbulence keeps dissipating kinematic energy from the system. When one is continuously injecting energy to
 the system, such that the energy injection rate is comparable to or larger than
 the energy dissipation rate of the viralised turbulence in the object, the object will be
``supported'' against collapse \footnote{When the external
energy injection rate is much larger than the energy dissipation rate of the
virialised turbulence, the object would be disrupted by turbulence cascade.}.
When the energy injection rate is not able to compensate for the internal energy dissipation, the object should collapse, such
that the gravitational energy released during this collapse would be able to
compensate for the additional energy dissipated by the viralised turbulence. In
contrast to the case of Jeans instability where \emph{pressure} plays a crucial rule, to
decided if a turbulent object will collapse, one need to come up
with an \emph{energy-based criterion}, which we derive in the next section.

Turbulence in astrophysical systems can be either supersonic or subsonic. For
subsonic turbulence, turbulent motion would lead to energy cascade, but the gas
compression from the turbulent motion is in general insignificant. In this case,
one would expect gravitational instability to develop gradually, perhaps
limited by a typical scale. Fragments developed from the
instability can still have different masses provided that they have different
ages. Yet, one still expect to observe a limit below which the instability can
not grow.

 When
the turbulence is supersonic, such as the case of the Milky Way interstellar
medium, turbulence itself creates density fluctuations, and the subsequent
interplay between turbulence and gravity determines the subsequent growth of the
perturbations created by turbulence.
In the supersonic case, one does not expect to observe the same limiting scale for gravitational instability to grow. The
instability can grow over a variety of scales provided that the initial density
fluctuations are large enough.  However, the critical condition for the density fluctuations to grow should still be determined by the
properties of the ambient turbulent flow.

The central part of our formalism is the ``condition for quasi-isolated
gravitational collapse'', as it is the theoretical boundary that separates the
dense, collapsing phase from an ambient diffuse phase of a turbulent medium.
In Sec. \ref{sec:formalism} we derive the condition for the quasi-isolated
gravitational collapse. When applied to a medium with subsonic
turbulence, this condition allows us to determine the critical condition
for a perturbation to grow (stability criterion). When applied to a medium where
the turbulence is supersonic or the density enhancements are pre-existing, the
condition allows us to determine if the structures are sufficiently condensed,
such that they would ignore the energy cascade from the ambient medium and would
evolve on their own.

\section{The formalism}\label{sec:formalism}
We consider the evolution of a dense object in a turbulent ambient medium. 
All the quantities have been listed in in Table \ref{table:1}. In this section
we will present our formulation, and an example is given in Sec.
\ref{sec:expalin}. The object has mass $m$ and size $l$.
The ambient medium has an almost-uniform mean density $\rho_{\rm medium}$.
%

The energy dissipation rate \footnote{Here, the energy dissipation rate is
obtained by averaging over a
few crossing times. } of the turbulent medium can be expressed as
\begin{equation}\label{eq:diss} 
   \epsilon_{\rm cascade} \approx  \eta \times  \frac{U_l^3}{l} \;
   \approx {\rm Constant},
\end{equation}
where $U_l$ is the velocity at scale
$l$, $\eta$ is the efficiency of turbulence dissipation, and is around $0.2$
\citep{2004RvMP...76..125M}.
When the injection scale of the turbulence is much larger than the scale of
interest, $ \epsilon_{\rm cascade}$ is roughly a constant, and is independent
on the scale (and thus holds for any given scale, see e.g. the energy
dissipation law, and \citet{2007ApJ...665..416K} for the supersonic
case).

Then we consider the effect of gravity on such an object, and temporarily
neglects the effect of external turbulence. Because of gravity, the object
would be dominated by a turbulent motion that is viralised, such that the
internal velocity dispersion is $U_{{\rm vir}, l} \approx \sqrt{G m/l}$.
The energy dissipation rate of such a turbulent, self-gravitating system is
\begin{equation}\label{eq:diss:grav}
 \epsilon_{\rm vir} \approx \eta \times \frac{U_{{\rm vir}, l}^3}{l} 
 = G^{3/2}\; m^{3/2} \;l^{-5/2}  \times \eta   \;.
\end{equation} 

We combine the
 above-mentioned results to propose our criterion. At the boundary of an object,
 the density of the gas that belongs to the object is comparable to the density
 of the gas of the surrounding. When $ \epsilon_{\rm vir}< \epsilon_{\rm
 cascade}$, energy injection from the ambient medium exceeds by much the energy dissipation
 of the turbulence inside the object were the turbulence is viralised. In this
case, self-gravity has an almost negligible effect on the system, and the system
is ``supported'' against collapse by energy from turbulence cascade.
When $ \epsilon_{\rm vir}> \epsilon_{\rm cascade}$, the {\it internal }energy
dissipation rate of the turbulent motion driven by self-gravity exceeds by much
the  {\it external} energy cascade from the ambient medium, and the system would
neglect the energy flux from the external cascade and would collapse on its own.
 Thus we propose a criterion for {\it quasi-isolated gravitational collapse}:
 \begin{equation}\label{eq:crit}
  \epsilon_{\rm vir} \geq  \epsilon_{\rm cascade}\;.
\end{equation}
When
 this is fulfilled, energy contribution from
 external turbulence cascade would not be able to influence the evolution
 significantly. The energy dissipation from the internal, viralised turbulence
 dominates the kinetic energy budget of the system, and the system would undergo
gravitational collapse. \footnote{
  Our distinction between the ``object'' and the ``ambient medium'' shares a
 common spirit with the distinction between ``two-stages'' in
 \citet{2012ApJ...750...13C}, 
 ``gravoturbulent regime and predominantly turbulent'' regime proposed in
 \citet{2015MNRAS.451.1056S}, and an observational correspondence would be the
 distinction between turbulence-type (t-type) and gravity-type (g-type)
 molecular clouds discussed in \citet{2016arXiv160304342L}.}
 The critical mass beyond which gravity dominates is (letting $\epsilon_{\rm
 grav} \approx \epsilon_{\rm cascade}$).



%

\begin{equation}\label{eq:mr}
m_{\rm crit} \approx G^{-1}  \epsilon_{\rm cascade}^{2/3}  \eta^{-2/3} 
l^{5/3}\;,
\end{equation}
where  $m_{\rm crit}$ is the critical mass and $l$ is the size. 
\begin{table*}
\begin{center}
\begin{tabular}{ |c|c|c| } 
 \hline
 Symbol & Quantity  & Unit \\ 
\hline
$l$ & size, scale &  $L$\\
$m$ & mass & $M$\\
$\rho$ & density & $M / L^3$\\
$G$ & gravitational constant &  $M^{-1} L^{3} T^{-2}$\\
 $U_l$ & velocity at scale $l$ & $L/T$\\
$U_{{\rm vir},l}$ & velocity dispersion of the viralised turbulence &
$L/T$\\

 $\epsilon_{\rm cascade}$ & energy dissipation rate of external
 turbulence \newline of order $\eta U_l^3 / l $ & $L^2 T^{-3}$ \\
  $\epsilon_{\rm vir}$ & energy dissipation rate the virialised 
 turbulence & $L^2 T^{-3}$\\
  $\epsilon_{\rm collapse}$ & energy injection from gravitational collapse& $L^2
  T^{-3}$\\
 $\eta\approx 0.2$ & efficiency of turbulent energy dissipation & $1$\\  
 $m_{\rm crit}$  &critical mass for turbulent gravitational instability & $M$ \\
  $l_{\rm crit}$ & critical scale for turbulent gravitational instability &
  $L$\\ $M$ & mass of the star cluster-forming clumps  & $M$\\
$R$ & radius of the star cluster-forming clumps & $L$\\
  \\

  \hline

\end{tabular}
\end{center}
\caption{\label{table:1} List of definitions of mathematical symbols. In the
right column, $L$ stands for scale, $T$ stands for time, and $M$ stands for
mass.}
\end{table*}

\subsection{Roles of turbulence and gravitational
contraction}\label{sec:expalin}

In our formalism, turbulent motion prevails throughout the region. However, the
function of turbulence is different at different regimes. 
Outside the object, an almost universal turbulence provides supports against
collapse. Inside the object, the turbulence is driven by a combination of
cascade from the large scale and gravitational collapse (e.g. turbulence driven
by accretion \citep{2010A&A...520A..17K,2010ApJ...712..294E}, and the energy
from gravitational collapse would dominate the energy budget
(such that $\epsilon_{\rm vir} - \epsilon_{\rm cascade} \approx \epsilon_{\rm
collapse}$ where $\epsilon_{\rm collapse}$ is the energy injection from
gravitational collapse, and $\epsilon_{\rm vir} \approx \epsilon_{\rm
collapse}$). As for the evolution of the object, it is likely that how fast the
object can collapse is determined by the ability of the turbulent motion to
remove the kinetic energy (See an analytical
 model from  \citet{2015ApJ...804...44M} and discussions therein).

\section{Gravitational instability in a turbulent medium}
The development of gravitational instability in a turbulent medium
has been studied
\citep{1951RSPSA.210...26C,1952Natur.170.1030P}.
However, these models typically assume a uniform velocity dispersion for the gas. Modern studies of
turbulence reveal that it is a multi-scaled process
\citep{1995tlan.book.....F}, and thus our  Eq. \ref{eq:crit} is  much more
accurate than assuming a constant velocity dispersion.

We consider the growth a perturbation on scale $l$.  The energy injection from
the turbulence cascade of the ambient medium is $\epsilon_{\rm cascade}$, and the total energy dissipation from the viralised 
turbulence were gravity to dominate is $\epsilon_{\rm vir}$. For the
instability to grow, one requires $\epsilon_{\rm vir} > \epsilon_{\rm cascade}$, which is identical to our criterion
of quasi-isolated gravitational collapse. Thus, Eq. \ref{eq:mr} still holds, and
at the onset of the gravitational instability, $\rho_{\rm medium} \approx
\rho_{\rm object} \approx \rho$. The {\it critical length scale on which the
instability can develop} is (where we have used Eq. \ref{eq:mr} and $m \approx \rho\, l_{\rm crit}^3$)
\begin{equation}\label{eq:l}
l_{\rm crit} \approx \epsilon_{\rm cascade}^{1/2} \eta^{-1/2} G^{-3/4}
\rho^{-3/4}\;.
\end{equation}
and the critical mass 
\begin{equation}\label{eq:mt}
m_{\rm crit} \approx G^{-1}  \epsilon_{\rm cascade}^{2/3} \eta ^{-2/3} 
l^{5/3} = \epsilon_{\rm cascade}^{3/2} \eta^{-3/2} G^{-9/4} \rho^{-5/4}\;,
\end{equation}
where $\epsilon_{\rm cascade}$ is energy dissipation rate of the ambient
turbulence, $\rho$ is the density and $G$ is the gravitational constant.

We expect the turbulent gravitational instability to develop when the medium is
supported by a turbulence that is subsonic. In such a case, Eq.
\ref{eq:l} predicts the critical length for the instability
to develop. In the supersonic case, because
turbulence also creates density fluctuations, one can obtain structures that
are much smaller than the scale predicted by Eq. \ref{eq:l}.

\section{Observational test}
One example to consider is the evolution of star cluster-forming clumps in 
molecular clouds.
Clumps (the definition can be found in \citet{2000prpl.conf...97W}) are
condensations of dense gas.
They are thought to be the progenitors of star clusters. Here, star
cluster-forming clumps are our objects of interest, and gas in the molecular
clouds servers as the ambient medium.

The Milky Way molecular clouds are turbulent. Observationally, molecular clouds
follow the Larson's relation \citep{1981MNRAS.194..809L} $\sigma_{\rm v} \sim l^{\beta}$ where the scaling index $\beta$ is close to
$1/3$ (the observed values varies from 0.38 (measured by
\citet{1981MNRAS.194..809L}) to 0.5, e.g. \citep{2011ApJ...740..120R}, and this might be dependent on the techniques
used to derive the scaling index).
The dissipation rate of  turbulence is
$\sigma_{\rm v}^3 / l$ (where $\sigma_{\rm v}$ is the velocity dispersion and
$l$ is the size). When $\beta = 0.33$, the dissipation rate is
completely independent on the scale. The observations thus indicate an
almost-uniform energy dissipation rate of turbulence in Milky Way molecular
clouds.
 
Star cluster-forming clumps (sometimes simply called ``clumps'') are dense gas
condensations in molecular clouds.
\citet{2015arXiv151200334P} pointed out the resemblance
between the mass-size relation of the clumps $M_{\rm clump}/ M_{\odot}
= 2500 (R/{\rm pc}  )^{1.67\pm 0.01}$ and
 the mass-size relation of embedded clusters
(\citet{2003ARA&A..41...57L}, where $M_{\rm cluster}/ M_{\odot} = 359 (R/{\rm
pc} )^{1.71\pm 0.07}$). One attractive possibility proposed by these authors is
that these clumps would collapse and form individual star clusters. If this is the case,
 the two above-mentioned mass-size relations should share a common origin, which
we would now explore.

We interpret the clumps as objects that undergo quasi-isolated gravitational
collapse in the molecular ambient medium. This will simultaneously explain why
these clumps would collapse to form individual star clusters, as well as the observed mass-size relation. 
For clumps to be gravitationally isolated, our formalism requires the sizes and
the masses to obey Eq.
\ref{eq:mr}, which is
\begin{equation}
M \sim R^{5/3} \sim R^{1.67}\;.
\end{equation}
The derived scaling index matches {\it exactly} with that of the observed
mass-size relation of the clumps (where the scaling index is $1.67\pm 0.01$) as well as
that of the embedded star clusters (where the scaling index is $1.71\pm 0.07$).
One can also derive the normalisation. We first estimate the expected turbulence
energy dissipation rate. Using Eq. \ref{eq:diss} and put $\sigma_v \approx
1\;\rm km/s$, $l \approx 1\;\rm pc$ \citep{1981MNRAS.194..809L}, we derive an energy dissipate
rate of $3.3\times 10^{-4}\;\rm \; cm^{2} \; s^{-3} =  3.3\times 10^{-4}\;\rm \;
erg \; s^{-1} \; g^{-1} $ ($1 \rm erg  = 1 \rm g\; cm^2\; s^{-2}$).
This gives us a mass-size relation $M_{\rm clump}/ M_{\odot} = 700 M_{\odot}
(R/{\rm pc})^{1.67}$, which agrees with the measured relation of embedded star clusters (where the
normalisation is 357 $M_{\odot}$, \citet{2003ARA&A..41...57L}), and does not
contradict the observed mass-size relation of the clumps
(where the normalisation is 2500 $M_{\odot}$,
\citet{2015arXiv151200334P,2013MNRAS.435..400U}). There are some differences. As
\citet{2015arXiv151200334P} have emphasised, the differences might arise simply
because of the different ways to define the radii.

Out result thus offers an explanation to the observed mass-size relation, and
shed light on the nature of these objects -- the so-called cluster-forming
clumps are objects that undergo quasi-isolated gravitational collapse, and their
sizes are the boundaries beyond which gravity from the central object ceases to
dominate \footnote{Inside the clump boundaries, $ \epsilon_{\rm vir}$  dominates, and outside the clump
boundaries, $\epsilon_{\rm cascade}$, dominates. One thus need to require that the
turbulence energy dissipation rate increases with decreasing radii.
The
dissipation inside the clump can be easily derived assuming some density profiles and with our Eq.
\ref{eq:diss:grav}. When the  inside-out density gradient is steeper than $\rho
\sim r^{-1/3}$, the energy dissipation rate increases as one moves inward, and
at regions inside the clumps, the external turbulence cascade does not
contribute much to the energy budget.
This condition is easily fulfilled for the majority of the observed clumps (where,
typically, $\rho \sim r^{-2}$, See e.g.
\citet{2016A&A...585A.149W}). }. Based on
this interpretation, we expect the clump evolution to be dynamically detached
from the ambient medium, and  thus they should collapse individually and form
star clusters.
This also explains the resemblance between the mass-size relation of the
clumps and that of the embedded star clusters \footnote{Our interpretation
should be distinguished from the interpretation of
\citet{2013ApJ...779..185K}, where the mass-size relation originates from the
Larson relation, which is an empirical relation derived from molecular cloud
observations. We agree that the Larson relation plays an important role, and
yet, in our formalism, the mass-size relation is rather a consequence of an
almost-uniform turbulence energy dissipation of the Milky Way molecular ISM.}.

One clarification should be made: the clumps are self-gravitating, and are
dynamically detached from their ambient environment, and this does not
imply that the interactions with the ambient medium are completely negligible.
We still expect the clumps to interchange matter with the ambient medium. {\it In short, the object is not completely isolated, but
is quasi-isolated, in that the energy exchange between the object and the
ambient medium is not sufficient to influence the gravitational collapse
significantly.}


For the molecular clouds in the Milky Way, because supersonic turbulence also
creates density fluctuations upon which gravity acts, we do not expect them to
respect the stability criterion (Eq. \ref{eq:l}), but they should still respect
the criterion for quasi-isolated gravitational collapse (Eq. \ref{eq:mr}). Only when
the turbulence is subsonic (and is close to be incompressible) do we expect our Eq.
\ref{eq:l} to predict the typical mass of the fragments that develop from
turbulent gravitational instability.

The importance of magnetic field in molecular clouds has been increasing
recognised \citep{2014prpl.conf..101L}. Our formalism does not include the
magnetic field.
However, how magnetic fields evolve in such a turbulent medium is still not
clear. { Observationally, systematic magnetic field measurements
are only available for nearby molecular clouds \citep{1999ApJ...520..706C}, and
measurements of field strength in massive star-forming regions are still limited
to individual sources
\citep{2015ApJ...799...74P,2015Natur.520..518L,2014ApJ...792..116Z}.  In this
sense, a complete picture of the evolution of the field strength in
different objects is still missing.} One theoretical possibility is that the
magnetic fields are maintained by turbulent motion and Galactic shear, and
the very process of field amplification by turbulence would lead to turbulent
magnetic reconnection, which might enable a relatively fast removal of
the magnetic flux \citep{2005PhR...417....1B,1993IAUS..157..427L}. If this is
the case, the effect of magnetic field would be secondary as compared to turbulence. But the
issue is still unsettled, and more investigations are needed.
\section{Conclusions}

We derive an analytical criterion for an object to undergo quasi-isolated
gravitational collapse in a turbulent medium. Different from the previous
treatments assuming constant velocity dispersions for the turbulence
\citep{1951RSPSA.210...26C,1952Natur.170.1030P}, we describe the multi-scaled
structure of the turbulent flow.
Our main results include a criterion for quasi-isolated gravitational
collapse and a condition for turbulent gravitational instability.

The criterion of quasi-isolated gravitational collapse allows one to decide if
an object is dynamically detached from the ambient turbulent flow. The critical
mass is linked to the size of the object $l$ and energy dissipation rate of the
ambient medium $\epsilon_{\rm cascade}$ by
 \begin{equation}
m_{\rm crit} \approx G^{-1}  \epsilon_{\rm cascade}^{2/3} \eta ^{-2/3} 
l^{5/3}  \;,\nonumber
\end{equation}
where $\eta \approx 0.2$ is an efficiency factor.
This result is applicable to both supersonic and subsonic flows.

We also derive a condition for turbulent gravitational instability to develop:
the critical scale $l_{\rm crit}$ is determined by the density of the medium
$\rho$ and the energy dissipation:
\begin{equation}
l_{\rm crit} \approx \epsilon_{\rm cascade}^{1/2} \eta^{-1/2} G^{-3/4}
\rho^{-3/4}\;.\nonumber
\end{equation}
This formula is applicable to
the subsonic case where we expect the instability to develop gradually from a
medium of a almost-uniform density. 

 Our criterion for quasi-isolated gravitational collapse  explains the observed
 mass-size relation of the star cluster-forming clumps
 $M_{\rm clump}\sim r^{1.67\pm 0.01}$, and thus supports a scenario that these
 objects {are dynamically detached from their environments,} and are 
 undergoing quasi-isolated gravitational collapse.\\

 {\noindent \it Note added in proof: A series of efforts have been made by
 \citet{2012A&A...545A.147H,2016A&A...591A..30L,2016A&A...591A..31L}, where they
 consider the interplay between turbulence dissipate and accretion, and derived an analytical mass-size relation $m\sim r^2$. Their picture shares many similarities with ours. However, to derive the mass-size relation, they had to assume Larson's relation, and in our case, the mass-size relation is a direct consequence of the universal turbulence dissipation in the ambient medium. Future observations of fragmentation in different environments are required to identify the key physical mechanism that lead to the mass-size relation. }

{\it Acknowledgements} 
Guang-Xing Li is supported by the Deutsche Forschungsgemeinschaft (DFG)
priority program 1573 ISM-SPP.








\appendix

\bibliography{paper}


\bsp	
\label{lastpage}
\end{document}